%% file: paper.tex
\documentclass[preprint]{ptephy}
\preprintnumber{YITP-15-79, KUNS2576}
\usepackage{boites}
\usepackage{graphicx}
\usepackage{dcolumn}
\usepackage{bm}
\usepackage{braket}
\begin{document}

\title{$\alpha$-cluster excited states in $^{32}$S}

\author{\name{\fname{Yuta} \surname{Yoshida}}{1\ast}, \name{\fname{Yoshiko} \surname{Kanada-En'yo}}{2}, and \name{\fname{Fumiharu} \surname{Kobayashi}}{3} }


\address{
\affil{1}{Yukawa Institute for Theoretical Physics, Kyoto University, Kyoto 606-8502, Japan}
\affil{2}{Department of Physics, Kyoto University, Kyoto 606-8502, Japan}
\affil{3}{Department of Physics, Niigata University, Niigata 950-2181, Japan}
\email{yyuta@yukawa.kyoto-u.ac.jp}
}

\begin{abstract}
$\alpha$-cluster excited states in $^{32}$S are investigated with an extended $^{28}$Si+$\alpha$ cluster model, in which the $^{28}$Si core deformation and rotation, and the $\alpha$-cluster breaking are incorporated.
In the GCM calculation with the extended $^{28}$Si+$\alpha$ cluster model, the $\alpha$-cluster excited states are obtained near the $^{28}$Si+$\alpha$ threshold energy.
The $^{28}$Si core deformation and rotation effects, and also the $\alpha$-clusters breaking in the $^{28}$Si+$\alpha$ system are discussed.
It is found that the rotation of the oblately deformed $^{28}$Si core gives a significant effect to the $\alpha$-cluster excited states whereas the $\alpha$-cluster breaking gives only a minor effect.
\end{abstract}

\subjectindex{D11}

\maketitle


\input{INTRODUCTION}
\input{FRAMEWORK}
\input{RESULTS}
\input{DISCUSSION}

\input{CONCLUSION}
\input{ACKNOWLEDGEMENTS}

\input{biblography}

\end{document}

%% file: INTRODUCTION.tex

\section{INTRODUCTION}
\label{sec:INTRO}

Cluster structure is one of the important aspects in nuclear system, in particular, in light nuclei.
The $\alpha$-cluster excited states having a spatially developed $\alpha$ cluster around a core nucleus have been known in $Z=N$ nuclei \cite{supp80,Ohkubo-rev,Oertzen-rev,Freer2007,AMDsupp-rev,Horiuchi-rev}, 
and also in unstable nuclei \cite{FREER,Ito:2003px,Curtis:2004wr,Soic:2003yg,Suhara:2010ww,Descouvemont:1985zz,Gai:1983zz,Gai:1987zz,Furutachi:2007vz,Fu:2008zzf,Johnson:2009kj,oertzen-o18,Curtis:2002mg,Ashwood:2006sb,Scholz:1972zz,Descouvemont:1988zz,Kimura:2007kz,Rogachev:2001ti}. 
Typical example of the $\alpha$-cluster excited states in $Z=N$ nuclei are the $^{16}{\rm O}+\alpha$ cluster states in $^{20}$Ne and $^{12}{\rm C}+\alpha$ cluster states in $^{16}$O \cite{Horiuchi1968,Hiura1969,wildermuth72,Nemoto1972,Matsuse73,Tanabe1974,Suzuki1976,Libert80}. 
The $\alpha$-cluster excited states are also suggested in the heavier mass nuclei such as $^{44}$Ti and $^{40}$Ca  
\cite{Michel1988,Wada1988,Ohkubo1988Ti,Ohkubo1988Ca,Merchant1989,Reidemeister1990,Yamaya1990a,Yamaya1990b,Yamaya1993Lett,Yamaya1993Rev,Sakuda1994,Yamaya1996,Yamaya1998suppl,Saukuda1998suppl,Kimura2006}. 

Candidates for the $\alpha$-cluster excited states in $^{32}$S have been reported in the $^{28}$Si($^{6}$Li,d)$^{32}$S ($\alpha$ transfer) reaction by Tanabe {\it et al.}\cite{Tanabe1981} in the 1980s.
A couple of states observed in $10\sim 15$ MeV region may correspond to the $\alpha$-cluster excited states.
Recently, in the experiments of the $^{28}{\rm Si}+\alpha$ elastic-scattering reaction, L\"onnroth {\it et al.} observed many resonances above the $^{28}{\rm Si}+\alpha$ threshold energy, and interpreted them as fragmentation of $\alpha$-cluster excited band starting from the bandhead energy $E_x=10.9\pm 0.5$ MeV, a few MeV higher energy than the $^{28}{\rm Si}+\alpha$ threshold \cite{Lonnroth2011}.
Another experiment for the $\alpha$-cluster excited states in $^{32}$S is the inelastic scatterings on $^{32}{\rm S}$ by Itoh {\it et al.} \cite{Itoh2012}.
They observed excited states near the $^{28}{\rm Si}+\alpha$ threshold energy are considered to be candidates for $\alpha$-cluster excited bands with the bandhead energies $E_x=6.6$ and $7.9$ MeV.
To understand the $\alpha$-cluster excited states in $^{32}$S, theoretical studies are now requested.

In a history of theoretical studies of cluster structures in the $p$-shell and $sd$-shell regions,
multi-$\alpha$ models using the Brink-Bloch $\alpha$-cluster wave functions \cite{Brink1970} have been applied to $Z=N=$ even nuclei. 
With the multi-$\alpha$ models, systematic calculations of 3-dimensional $\alpha$-cluster configurations were performed from $^{16}$O to $^{44}$Ti \cite{Zhang1994}.
For $^{28}$Si, the 7$\alpha$-cluster model was used to discuss the shape coexistence of the oblate and prolate states \cite{Bauhoff1982}.
The multi-$\alpha$ model was also used for $^{20}$Ne to take into account the $^{16}$O core structure change in $^{16}$O+$\alpha$ cluster states in $^{20}$Ne \cite{Nemoto1975}.
However, in these studies with the multi-$\alpha$ models,
constituent $\alpha$ clusters are assumed  to be the ideal $0s$-closed configuration,
and therefore the contribution of the spin-orbit interaction is completely omitted even though it is significant in mid-shell nuclei. 
In other words, $\alpha$ clusters in nuclei should be more or less broken from the ideal configuration to gain the spin-orbit interaction.
To take into account the $\alpha$-cluster breaking and the contribution the spin-orbit interaction,
an extension of cluster models has been done in the study of the $^{16}$O+$\alpha$ cluster states in $^{20}$Ne \cite{Itagaki2011}.
Cluster structures in $sd$-shell nuclei were also investigated by the antisymmetrized molecular dynamics (AMD) \cite{AMDsupp-rev},
in which the existence of clusters is not assumed but the formation and breaking of cluster structures are automatically described in the model.
In the AMD calculation for $^{28}$Si, the oblately deformed state with a $7\alpha$-like configuration was obtained for the $^{28}$Si ground state consistently with the $7\alpha$-cluster model calculation \cite{Bauhoff1982}, however, it was shown that the oblate ground state is different from the ideal $7\alpha$ configuration but it contains the significant cluster breaking because of the spin-orbit interaction \cite{Enyo2005}.
In the systematic studies with the AMD by Taniguchi {\it et al.}, the $\alpha$-cluster excited states were suggested in various $sd$-shell nuclei \cite{Taniguchi2007,Taniguchi2009}.
In these studies, the existence of clusters are not assumed {\it a priori}, but core deformation and the $\alpha$-cluster breaking are taken into account in the AMD framework. However, the rotation of the core in the $\alpha$-cluster excited states is not sufficiently considered.

Our aim in this paper is to theoretically investigate the $\alpha$-cluster excited states in $^{32}$S. The question to be answered is whether the $\alpha$-cluster band appears near the $^{28}$Si+$\alpha$ threshold energy.
If the case, we are going to predict its properties such as the bandhead energy, the level spacing (the rotational constant), and the $\alpha$-decay width.
We also intend to clarify the core deformation and rotation effects as well as the $\alpha$-cluster breaking effect in the $\alpha$-cluster excited states.
In $\alpha$-cluster excited states in the $sd$-shell region, the core deformation may occur, and the rotation of the deformed core could play an important role.
Moreover, an $\alpha$ cluster at the nuclear surface can be dissociated because of the spin-orbit potential.
To incorporate the core deformation and rotation as well as the $\alpha$-cluster breaking, we construct a new extended cluster model for the $^{28}{\rm Si}+\alpha$ system by extending the conventional cluster model, which relies on the inert cluster assumption.
We apply the method and investigate the properties of $\alpha$-cluster excited states in $^{32}$S.

The contents of this paper are as follows.
In Sec. \ref{sec:framework}, we explain the formulation of the extended  $^{28}{\rm Si}+\alpha$-cluster model. 
We show the calculated results in Sec. \ref{sec:results},
and discuss the $^{28}$Si core structure and the $\alpha$-cluster breaking effect in the $\alpha$-cluster excited states in $^{32}$S in Sec. \ref{sec:discussion}.
Finally, a summary and an outlook are given in Sec. \ref{sec:conclusion}.

%% file: FRAMEWORK.tex

\section{FRAMEWORK}
\label{sec:framework}

To investigate $\alpha$-cluster excited states in $^{32}$S,
we construct the extended cluster model for the $^{28}$Si+$\alpha$ system 
to take into account the $^{28}$Si core deformation and rotation, and the $\alpha$-cluster breaking.
In this section, we first explain the Brink-Bloch $\alpha$-cluster model (a conventional cluster model),
and then, we describe the formulation of the extended $^{28}$Si+$\alpha$-cluster model.

\subsection{Brink-Bloch $\alpha$-cluster model}

In the Brink-Bloch $\alpha$-cluster model \cite{Brink},
a $Z=N=2n$ nucleus is composed of $n\alpha$ clusters.
Each $\alpha$ cluster is described by the $(0s)^4$ harmonic oscillator (h.o.) configuration
localized around a certain position. 
The total $n\alpha$-cluster wave function ${\rm \Phi}_{n\alpha}$ of the $A=4n$-body system is 
written by the following antisymmetrized single-particle wave functions, 
\begin{eqnarray}
{\rm \Phi}_{n\alpha}({\bf R}_1,\cdots,{\bf R}_n)
   &=& 
{\mathcal A} [\Phi_{\alpha}({\bf R}_1) \cdots  \Phi_{\alpha}({\bf R}_n)], \\
\Phi_{\alpha}({\bf R}_i) 
   &=&
       \varphi_{\uparrow p}({\bf R}_i)\varphi_{\downarrow p}({\bf R}_i)
       \varphi_{\uparrow n}({\bf R}_i)\varphi_{\downarrow n}({\bf R}_i), \\
\varphi_{\sigma}({\bf R}_i)
   &=& \left(\frac{2\nu}{\pi}\right)^{\frac{3}{4}}
       \exp \left[ -\nu\left({\bf r}-{\bf R}_i\right)^2 \right]
       \chi_{\sigma}
       \tau_{\sigma},
\end{eqnarray}
where $\mathcal{A}$ is the antisymmetrizing operator for all nucleons, 
${\bf R}_i$ is the center of the $i$th ${\alpha}$-cluster ($i=1,\cdots,n$), 
$\chi_\sigma$ and  $\tau_\sigma$ are the spin and isospin parts of the single-particle wave function, 
respectively, and $\nu$ is the width parameter.

\subsection{Extended cluster model for $\alpha$-cluster breaking}

For description of the $\alpha$-cluster breaking due to the spin-orbit potential from a core,
we apply the method proposed by Itagaki {\it et al.} \cite{Itagaki2011}. 
In this method, an $\alpha$-cluster breaking is incorporated  
by adding a spin-dependent imaginary part to the Gaussian centers of 
single-particle wave functions so as to gain the spin-orbit potential,
\begin{eqnarray}
 \Phi_{\alpha'}({\bf R},\lambda_{\alpha}) 
   &=& \mathcal{A} \left[  
       \varphi_{\uparrow p}({\bf Z}_1)\varphi_{\downarrow p}({\bf Z}_2)
       \varphi_{\uparrow n}({\bf Z}_3)\varphi_{\downarrow n}({\bf Z}_4)
       \right],\\
{\bf Z}_j &=& {\bf R}+  
       i\lambda_{\alpha} \frac{({\bf e}_{spin,j}) \times (\hat{\bf R})}{\sqrt{\nu}},
\label{Zi}
\end{eqnarray}
where the parameter $\lambda_{\alpha}$ represents the degree of the $\alpha$-cluster breaking,  
and ${\bf e}_{spin,j}$ is the unit vector oriented to the intrinsic spin direction of the $j$th nucleon ($j=1, \cdots, 4$).
If $\lambda_{\alpha}$ is zero, this model becomes the conventional cluster model (Brink-Bloch $\alpha$-cluster model) and 
describes the intrinsic spin saturated state, where the expectation value of the spin-orbit potential vanishes.
When $\lambda_{\alpha}$ is positive, spin-up and spin-down nucleons in the $\alpha$ cluster obtain finite momenta 
with opposite directions so as to gain the spin-orbit potential.

\subsection{Extended cluster model for $^{28}$Si core}

\begin{figure}[h]
\begin{center}
\includegraphics[clip,width=12.0cm]{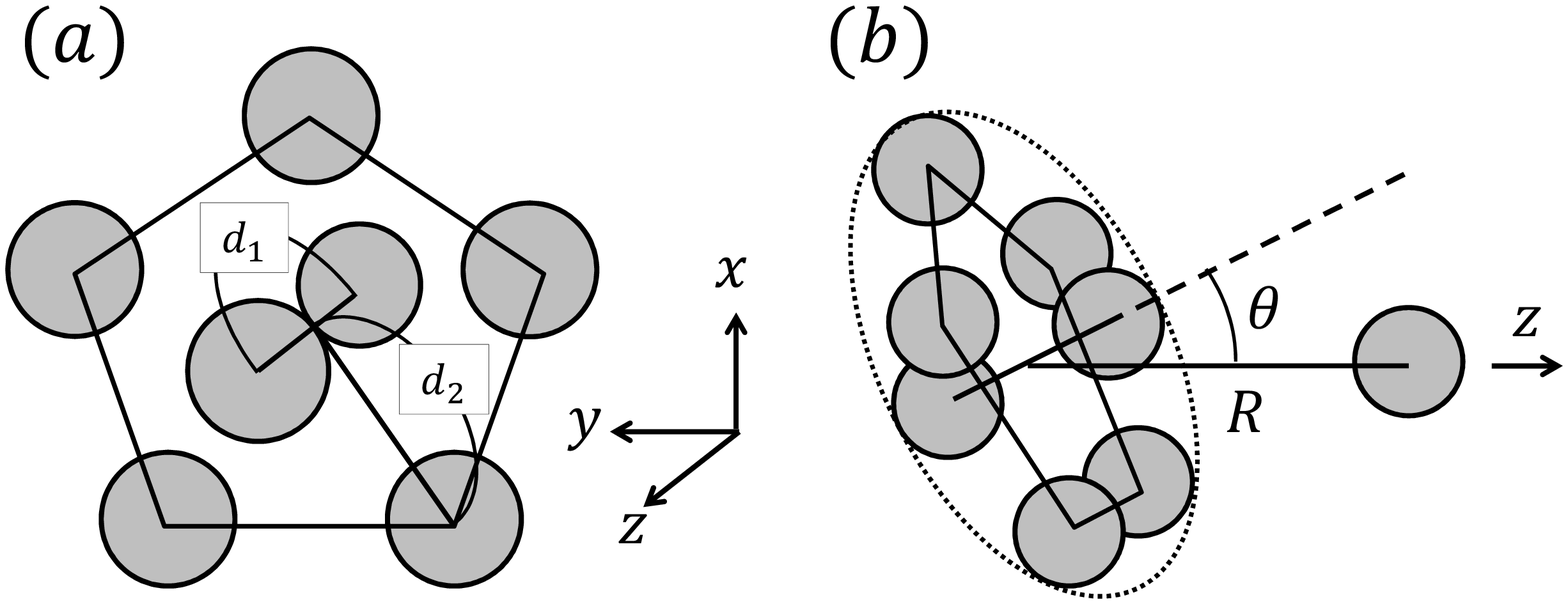}
\caption{\label{fig:si28_config}
Schematic figures for spatial configurations of Gaussian centers.
(a) The 7$\alpha$-cluster configuration with a pentagon shape 
for the $^{28}{\rm Si}$ core.
(b) The configuration for the present $^{28}{\rm Si}+{\alpha}$ cluster model.
}
\end{center}
\end{figure}

To describe the $^{28}$Si core structure, we adopt an extended $7\alpha$-cluster model 
where the parameter $\Lambda_c$ for the cluster breaking is incorporated to take into account the spin-orbit interaction effect.
The present extended $7\alpha$-cluster model is based on the study of $^{28}$Si with the Brink-Bloch $7\alpha$-cluster model \cite{Bauhoff1982} and 
that with the method of the AMD \cite{Enyo2005}.
Bauhoff {\it et al.} used the 7$\alpha$-cluster model with a pentagon configuration \cite{Bauhoff1982},
and succeeded to describe the oblate ground state and the $K^{\pi}=5^-$ rotational band
with the $D_{5h}$ symmetry of a pentagon configuration. 
The 7$\alpha$-cluster model wave function $\Phi_{7\alpha}$ forms a pentagon configuration as shown in 
Fig. \ref{fig:si28_config}(a), and is described as
\begin{eqnarray}
\Phi_{7\alpha}(d_1,d_2) 
   = \mathcal{A} \left[  
       \Phi_{\alpha}(\tfrac{1}{2}d_1{\bf e}_z) \Phi_{\alpha}(-\tfrac{1}{2}d_1{\bf e}_z)
       \prod_{k=1}^{5}  \hat{R}_z \left(\frac{2\pi}{5}k\right)  \Phi_{\alpha}(d_2{\bf e}_x)
       \right],
\end{eqnarray}
where $\hat{R}_z$ is the rotation operator around the $z$ axis, and $d_1$ and $d_2$ are the distance 
parameters for 7$\alpha$ cluster positions.

The pentagon configuration of the $7\alpha$-cluster structure of $^{28}$Si has been also supported by 
the AMD calculation where $\alpha$ clusters are not {\it a priori} assumed \cite{Enyo2005}. 
Differently from the Bauhoff's 7$\alpha$-cluster model,
the $^{28}$Si wave function obtained by the AMD for the ground state is not the ideal $7\alpha$-cluster wave function without the cluster breaking but it is a 28-body wave function with
a pentagon configuration of 7$\alpha$ clusters having the cluster breaking.

Based on the AMD result for $^{28}$Si, we construct an extended $7\alpha$-cluster model for the $^{28}$Si core 
by respecting the symmetry for the $2\pi/5$ rotation as follows,  
\begin{eqnarray}
&&\Phi_{^{28}{\rm Si}}(d_1,d_2,\Lambda_c) 
   = \mathcal{A} \left[  
       \Phi_{\alpha}(\tfrac{1}{2}d_1{\bf e}_z) \Phi_{\alpha}(-\tfrac{1}{2}d_1{\bf e}_z)
       \sum_{k=1}^{5} \hat{R}_z \left(\frac{2\pi}{5}k\right) \Phi_{\alpha'}(d_2{\bf e}_x,\Lambda_c)
       \right], \\
&&\Phi_{\alpha'}(d_2{\bf e}_x,\Lambda_c) 
   = \mathcal{A} \left[  
       \varphi_{\uparrow_y p}(d_2{\bf e}_x+i d_2 \Lambda_c{\bf e}_z)\varphi_{\downarrow_y p}(d_2{\bf e}_x-i d_2 \Lambda_c{\bf e}_z) \right. \nonumber \\
&&\left. \hspace{4cm}       \varphi_{\uparrow_y n}(d_2{\bf e}_x+i d_2 \Lambda_c{\bf e}_z)\varphi_{\downarrow_y n}(d_2{\bf e}_x-i d_2 \Lambda_c{\bf e}_z)
       \right].
\end{eqnarray}
Here, $\Phi_{\alpha'}(d_2{\bf e}_x,\Lambda_c)$ represents the wave function for a broken $\alpha$ cluster,
where $\uparrow_y$ and $\downarrow_y$ are the intrinsic spin of the $y$ direction,
and nucleon momenta takes the $z$ direction.
$\Lambda_c$ is the parameter for the cluster breaking in the 7$\alpha$-cluster model for the $^{28}$Si core and called the $7\alpha$-cluster breaking parameter in this paper.
In the case of the $d_1\rightarrow 0$ and $d_2\rightarrow 0$ limit,
this extended 7$\alpha$-cluster model wave function
$\Phi_{^{28}{\rm Si}} \left(d_1\rightarrow 0,d_2\rightarrow 0,\Lambda_c\right)$ describes
the $0d_{5/2}$ sub-shell closed configuration of the $jj$-coupling shell model at $\Lambda_c=1$ 
and the oblately deformed state at $\Lambda_c=0$.
Note that 5$\alpha$ clusters in the $^{28}$Si core are broken $\alpha$ clusters written by the 
previously mentioned method for the $\alpha$-cluster breaking proposed 
by Itagaki {\it et al}.
The concept of the present model for the $^{28}$Si core is similar to that 
of an extended $3\alpha$ cluster model for $^{12}$C proposed
by Suhara {\it et al.} \cite{Suhara2013}. 

In the present calculation, the parameters, $d_1$ and $d_2$,  for positions of 7$\alpha$-clusters are fixed to be
the optimized values, $d_1=0.20\ {\rm fm}$, $d_2=0.27\ {\rm fm}$, that give the minimum energy 
of $^{28}$Si in the 7$\alpha$-cluster model without 
the cluster breaking ($\Lambda_c=0$). 
Hereafter, we define the $^{28}$Si wave function 
with the fixed $d_1$ and $d_2$ values as
$\Phi_{^{28}{\rm Si}}(\Lambda_c) \equiv \Phi_{^{28}{\rm Si}}(d_1=0.20\ {\rm fm},d_2=0.27\ {\rm fm},\Lambda_c)$
parameterized by $\Lambda_c$.

\subsection{Extended cluster model for $^{28}{\rm Si}+{\alpha}$ system}

We construct the extended $^{28}{\rm Si}$+${\alpha}$-cluster model to take into account the $^{28}$Si core deformation and rotation, and the $\alpha$-cluster breaking. 
We set the $^{28}$Si core and the $\alpha$ cluster at the 
inter-cluster distance $R$, and perform the generator coordinate method (GCM)\cite{GCM} by treating $R$ as the generator coordinate.
The $\alpha$ cluster is parameterized by the $\alpha$-cluster breaking parameter $\lambda_\alpha$,
whereas the $^{28}$Si core is specified by the $7\alpha$-cluster breaking parameter $\Lambda_c$ 
which changes the $^{28}$Si core deformation from the oblate state to the spherical one.
In addition to these parameters, $R$,  $\lambda_\alpha$, and $\Lambda_c$, 
we consider the angle parameter $\theta$ 
to specify the orientation of the oblately deformed  $^{28}$Si core. 
We set the $\alpha$ cluster on the z-axis and define $\theta$ 
for the rotation of the $^{28}$Si core as shown in Fig.~\ref{fig:si28_config}(b). 
When $\theta=0^{\circ}$, the symmetric axis of the $^{28}$Si core agrees to the $z$ axis.

Then the $^{28}$Si+$\alpha$ wave function of the extended $^{28}$Si+$\alpha$-cluster model is written as
\begin{eqnarray}
 \Phi_{^{28}{\rm Si}+\alpha}(R,\theta,\lambda_{\alpha},\Lambda_c)
&=& \mathcal{A} \left[ 
     \Phi_{\alpha'}        \left(\frac{7}{8}R{\bf e}_z,\lambda_{\alpha}\right) 
     \Phi_{^{28}{\rm Si}} \left(-\frac{1}{8}R{\bf e}_z,\theta,\Lambda_c\right)
               \right],\\
\Phi_{^{28}{\rm Si}}({\bf R},\theta,\Lambda_c)&=&\hat{T}({\bf R})\hat{R}_y(\theta)\Phi_{^{28}{\rm Si}}(\Lambda_c),
\end{eqnarray}
where $\hat{T}({\bf R})$ is the translation operator and $\hat{R}_y(\theta)$ is the rotation operator around the $y$ axis.
$\Phi_{^{28}{\rm Si}}({\bf R},\theta,\Lambda_c)$ expresses the extended 7$\alpha$-cluster model $\Phi_{^{28}{\rm Si}}(\Lambda_c)$ rotated by the angle $\theta$ and shifted by ${\bf R}$. 
In the extended $^{28}{\rm Si}$+${\alpha}$-cluster model, the width parameter is chosen to be $\nu=0.16 {\rm fm}^{-2}$ so as to reproduce the $^{28}$Si radius with the sub-shell closed configuration.

\subsection{Parity and total-angular-momentum projection}
We project the $^{28}$Si+$\alpha$ wave function $\Phi_{^{28}{\rm Si}+\alpha}(R,\theta,\lambda_{\alpha},\Lambda_c)$
to the parity and total angular-momentum eigenstate, 
\begin{equation}
\Phi^{J^{\pm}}_{^{28}{\rm Si}+\alpha}(R,\theta,\lambda_{\alpha},\Lambda_{c})
= \hat{P}^{J}_{MK}\hat{P}^{\pm} \Phi_{^{28}{\rm Si}+\alpha}(R,\theta,\lambda_{\alpha},\Lambda_c),
\label{ang}
\end{equation}
where $\hat{P}^{\pm}$ and $\hat{P}^{J}_{MK}$ are
the parity and the total-angular-momentum projection operators, respectively.
In the present paper, we only take the $K=0$ component and omit the $K$-mixing for simplicity.

\subsection{Generator coordinate method (GCM)}
\label{FW:GCM}

To calculate energy levels of $\alpha$-cluster states in $^{32}$S,
we perform the GCM calculation by superposing the $^{28}{\rm Si}+{\alpha}$ wave function, 
\begin{equation}
\Psi^{J^{\pm}_n}_{^{28}{\rm Si}+\alpha}
  = \sum_{i} c^{(n)}_i 
\Phi^{J^{\pm}}_{^{28}{\rm Si}+\alpha}(R_i,\theta_i,\lambda_{\alpha}=0,\Lambda_{ci}),
\label{GCMwf}
\end{equation}
where coefficients $c^{(n)}_i$ are determined by diagonalizing the norm and Hamiltonian matrices. 
For the inter-cluster distance $R$, we superpose the wave functions with $R=1,2,\cdots,10$ fm.
For the $7\alpha$-cluster breaking parameter $\Lambda_c$ of the $^{28}$Si core, we take two points,
$\Lambda_c=0.38$ and $ 0.80$, which correspond to oblate and spherical local minimum states 
of the intrinsic energy of the $^{28}$Si core as described later.
For the rotation angle $\theta$ of the $^{28}$Si core, we take $\theta= 0^{\circ}, 30^{\circ}, 60^{\circ}, 90^{\circ}$ 
for the oblate core ($\Lambda_c=0.38$)
and take $\theta=0$ for the spherical core ($\Lambda_c=0.80$). 

In the present GCM calculation, we omit the $\alpha$-cluster breaking and fix $\lambda_\alpha=0$.
More details of the choice of the parameters are described later. 
We call the calculation with the full diagonalization of the norm and Hamiltonian matrices in the above-mentioned basis wave functions
with the parameters  $(R_i,\theta_i,\Lambda_{ci})$ ``full-GCM'' calculation.

\subsection{Frozen core GCM}
\label{FW:Frozen_core}

In the asymptotic region at a large inter-cluster distance $R$,
the $^{28}$Si core in the lowest $^{28}{\rm Si}+{\alpha}$ channel should be the ground state of an isolate $^{28}$Si: $^{28}{\rm Si}(0^+_{g.s.})$.
We also perform the GCM calculation for the $^{28}{\rm Si}(0^+_{g.s.})+\alpha$ within the frozen core approximation and compare the result with the previously explained full GCM calculation. We call this calculation ``frozen core GCM''.
In the present work, we express the frozen core wave function by the linear combination of the projected $^{28}{\rm Si}(0^+_{g.s.})+\alpha$ wave functions as follows. 

Let us first consider the adiabatic picture that the $^{28}{\rm Si}$ configuration is optimized at each state of a given inter-cluster distance $R$.
We define the $R$-fixed $^{28}{\rm Si}+\alpha$ wave function as
\begin{eqnarray}
\Phi^{J^{\pm}}_{^{28}{\rm Si}'+\alpha} (R) = 
\hat{P}^{J}_{MK}\hat{P}^{\pm} \mathcal{A}\left[ \Phi_{\alpha}(\tfrac{7}{8}R,\lambda_{\alpha})
 \sum_{k} a_k(R) \Phi_{^{28}{\rm Si}}(-\tfrac{1}{8}R,\theta_k,\Lambda_{ck}) \right],
\label{P_si+a(R)}
\end{eqnarray}
where
parameters $(\theta_k,\Lambda_{ck})=(0^{\circ},0.38), (30^{\circ},0.38), (60^{\circ},0.38), (90^{\circ},0.38),$ and $ (0^{\circ},0.80)$ are taken.
Here coefficients $a_k(R)$ are determined by diagonalizing the norm and Hamiltonian matrices for each inter-cluster distance $R$.
By taking an enough large inter-cluster distance $R_{\rm max}$, we determine the coefficients $a_k(R_{\rm max})$ in the asymptotic region, which approximately express the ground state configuration of the $^{28}{\rm Si}$ core. 
We take $R_{\rm max}=10$ fm in this paper.

Next, using the coefficients $a_k(R_{\rm max})$ determined at $R_{\rm max}$,
we define the $R$-fixed $^{28}{\rm Si}(0^+_{g.s.})$+$\alpha$ wave function with the frozen core (the $R$-fixed frozen core wave function),
\begin{eqnarray}
\Phi^{J^{\pm}}_{^{28}{\rm Si}(0^+_{g.s.})+\alpha} (R) = 
\hat{P}^{J}_{MK}\hat{P}^{\pm} \mathcal{A}\left[ \Phi_{\alpha}(\tfrac{7}{8}R,\lambda_{\alpha})
 \sum_{k} a_k(R_{\rm max}) \Phi_{^{28}{\rm Si}}(-\tfrac{1}{8}R,\theta_k,\Lambda_{ck}) \right],
\label{P_si0+a(R)}
\end{eqnarray}
Then, we perform the frozen core GCM calculation, that is, the GCM calculation of the $^{28}{\rm Si}(0^+_{g.s.})+\alpha$ cluster model 
by superposing the $^{28}{\rm Si}(0^+_{g.s.})$+$\alpha$ wave functions with different distance as
\begin{eqnarray}
\Psi^{J^{\pm}_n}_{^{28}{\rm Si}(0^+_{g.s.})+\alpha} 
  = \sum_{i} b_i^{(n)} \Phi^{J^{\pm}}_{^{28}{\rm Si}(0^+_{g.s.})+\alpha} (R_i),
\label{eq:wf_frozen_core}
\end{eqnarray}
where coefficients $b_k^{(n)}$ are determined by diagonalizing the norm and Hamiltonian matrices.

\subsection{$^{28}{\rm Si}(0^+_{g.s.})+\alpha$ amplitudes in $^{32}{\rm S}$ wave functions}
\label{FW:Overlaps}

In order to analyze the $\alpha$-cluster motion in $^{32}$S states obtained by the full-GCM and those by the frozen core GCM calculations,
we calculate the overlap between the $^{32}$S wave functions
with the $R$-fixed $^{28}{\rm Si}(0^+_{g.s.})$+$\alpha$ wave function to evaluate the $\alpha$-cluster component at $R$, 
\begin{eqnarray}
f^{J^{\pm}_n}_{^{28}{\rm Si}+\alpha}(R) &=& 
\left| \braket{\Phi^{J^{\pm}}_{^{28}{\rm Si}(0^+_{g.s.})+\alpha} (R)|\Psi^{J^{\pm}_n}_{^{28}{\rm Si}+\alpha} } \right|,
\label{eq:overlap_full}\\
f^{J^{\pm}_n}_{^{28}{\rm Si}(0^+_{g.s.})+\alpha}(R) &=& 
\left| \braket{\Phi^{J^{\pm}}_{^{28}{\rm Si}(0^+_{g.s.})+\alpha} (R)|\Psi^{J^{\pm}_n}_{^{28}{\rm Si}(0^+_{g.s.})+\alpha}} \right|.
\label{eq:overlap_frozen}
\end{eqnarray}

\subsection{Hamiltonian}
The Hamiltonian operator ($\hat{H}$) is
\begin{eqnarray}
 &&\hat{H} = \hat{T}+\hat{V}_{nuclear}+\hat{V}_{coulomb}-\hat{T}_G, \\
 &&\hat{V}_{nuclear} = \hat{V}_{c}+\hat{V}_{LS},
\end{eqnarray}
where $\hat{T}$ is the kinetic energy and $\hat{T}_G$ is the energy of the center-of-mass motion.
As for the effective nuclear force $\hat{V}_{nuclear}$, Volkov No.2 \cite{Volkov} is adopted as the central force $\hat{V}_{c}$ and the two-range Gaussian form of the spin-orbit term in the G3SR force\cite{G3SR} is used as the spin-orbit force $\hat{V}_{LS}$. 

The form of Volkov No.2 is given as 
\begin{eqnarray}
 \hat{V}_{c} = \sum_{i<j}^{A}\sum_{k=1}^{2} 
                     v_k \exp\left[ -\left(\frac{\hat{\bf r}_{ij}}{a_k}\right)^2 \right]
                     \left( W-MP_{\sigma\tau} \right), 
\end{eqnarray}
where $v_1=-60.65$ MeV, $v_2=61.14$ MeV, $a_1=1.80$ fm, $a_2=1.01$ fm. 
$M$ is the Majorana parameter that is an adjustable parameter.
In the present paper, we use $M=0.67$. 
With the Volkov force, reproductions of the binding energy of $^{32}$S 
and the $\alpha$-separation energy ($^{28}{\rm Si}+\alpha$ threshold) are
not satisfactory. We also use other $M$ values of the Volkov force 
to discuss the interaction dependence of the calculated results.

The spin-orbit force is given as 
\begin{eqnarray}
 \hat{V}_{LS} &=& \sum_{i<j}^{A}\sum_{k=1}^{2} 
                     u_k \exp\left[ -\left(\frac{\hat{\bf r}_{ij}}{b_k}\right)^2 \right]
                     P(^3O) \ \hat{\bf L}\cdot\hat{\bf S}, \\
 &&P(^3{\rm O}) = \frac{1+P_{\sigma}}{2}\frac{1+P_{\tau}}{2},
\end{eqnarray}
where $b_1=0.477$ fm, $b_2=0.600$ fm, and $P(^3{\rm O})$ is the triplet-odd projection operator.
We use the strength parameters $u_1=2000$ MeV and $u_2=-2000$ MeV which are the same as those used in Ref. \cite{Itagaki2011} for the $^{16}{\rm O}+\alpha$ system.
The Coulomb force $\hat{V}_{coulomb}$ is approximated by seven Gaussians.

%% file: RESULTS.tex

\section{RESULTS}
\label{sec:results}

\subsection{$^{28}$Si core structure in $^{28}$Si+$\alpha$ system}

\begin{figure}[h]
\begin{center}
\includegraphics[clip,width=9.0cm]{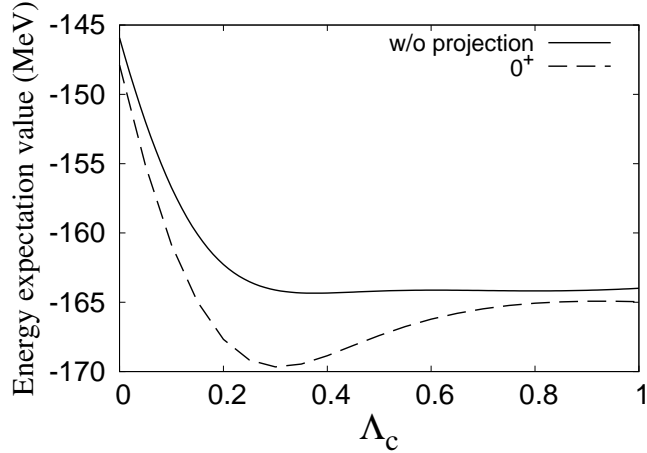}
\caption{\label{fig:Si28_lamc}
The energy expectation value of the isolate $^{28}$Si core.
The energy $E_{^{28}{\rm Si}}(\Lambda_c)$ before the parity and total-angular-momentum projection,
and the energy $E^{0^+}_{^{28}{\rm Si}}(\Lambda_c)$ after the projection are shown by solid and dashed lines, respectively.
The width parameter is taken to be $\nu=0.16 {\rm fm}^{-2}$.
}
\end{center}
\end{figure}

To discuss effects of the $7\alpha$-cluster breaking in the $^{28}$Si core because of the spin-orbit interaction,
we show, in Fig. \ref{fig:Si28_lamc}, the $\Lambda_c$ dependence of the energy of an isolate $^{28}$Si state 
before and after the parity and total-angular-momentum projection,
\begin{eqnarray}
 E_{^{28}{\rm Si}}(\Lambda_c)
 = \frac{\braket{ \Phi_{^{28}{\rm Si}}(\Lambda_c)|\hat{H}|\Phi_{^{28}{\rm Si}}(\Lambda_c) }}
        {\braket{ \Phi_{^{28}{\rm Si}}(\Lambda_c)        |\Phi_{^{28}{\rm Si}}(\Lambda_c) }},\\
 E^{0^+}_{^{28}{\rm Si}}(\Lambda_c)
 = \frac{\braket{ \Phi^{0^+}_{^{28}{\rm Si}}(\Lambda_c)|\hat{H}|\Phi^{0^+}_{^{28}{\rm Si}}(\Lambda_c) }}
        {\braket{ \Phi^{0^+}_{^{28}{\rm Si}}(\Lambda_c)        |\Phi^{0^+}_{^{28}{\rm Si}}(\Lambda_c) }},\\
\Phi^{0^+}_{^{28}{\rm Si}}(\Lambda_c) = \hat{P}^{J=0}_{M K=0} \hat{P}^+\Phi_{^{28}{\rm Si}}(\Lambda_c).
\end{eqnarray}
In the $\Lambda_c=0.3\sim 1.0$ region,
the $^{28}{\rm Si}$ system gains much energy of the spin-orbit interaction by the $7\alpha$-cluster breaking. 
In the energy curve of $E_{^{28}{\rm Si}}$ before the parity and total-angular-momentum projection,
there exist two energy minimums at $\Lambda_c=0.38$ and $\Lambda_c=0.80$ though the energy almost degenerates in this region.
We call these two minimums of $^{28}$Si the ``oblate-type ($\Lambda_c=0.38$)'' and ``spherical-type ($\Lambda_c=0.80$)'' states.
Here, the oblate-type state is different from the $\Lambda_c=0$ state that is the ideal state
with the $(200)^4(110)^4(020)^4$ configuration in terms of the $(n_x,n_y,n_z)$ notation of the h.o. shell-model basis in the $sd$ shell.
The energy of the oblate-type state at  $\Lambda_c=0.38$ is about $18$ MeV lower due to the $7\alpha$-cluster breaking than that 
of the $\Lambda_c=0$ state having no contribution of the spin-orbit interaction.
This result supports the AMD calculation of $^{28}$Si \cite{Enyo2005}
and indicates that the present method of the extended 7$\alpha$-cluster model is suitable to 
incorporate the significant energy gain of $^{28}$Si with the $7\alpha$-cluster breaking
in the oblately deformed $^{28}$Si.
In the $0^+$ projected $^{28}$Si energy, it is found that the oblate-type ($\Lambda_c=0.38$)
state gains further energy because of the restoration of the rotational symmetry.
The present result for the $^{28}$Si core indicates that the rotation of the oblately deformed state
can be an important degree of freedom of the $^{28}$Si core structure in the $^{28}{\rm Si}+{\alpha}$ system as well as
the $7\alpha$-cluster breaking due to the spin-orbit interaction.
 
\begin{figure}[h]
\begin{center}
\includegraphics[clip,width=9.0cm]{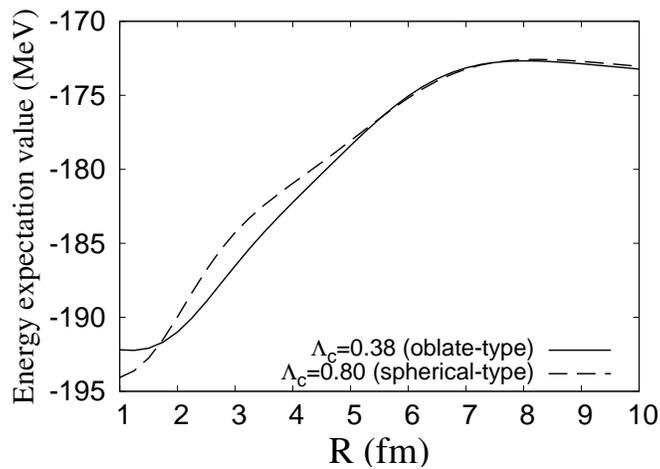}
\end{center}
\caption{\label{fig:proc_fig2}
Energy expectation value of $^{28}{\rm Si}+\alpha$ system ($E^+_{^{28}{\rm Si}+\alpha}(R,\theta=0^{\circ},\lambda_{\alpha}=0,\Lambda_c)$) as a function of the inter-cluster distance $R$. The parameter $\Lambda_c$ for the $^{28}$Si core structure is fixed to be $\Lambda_c=0.38$ (oblate type:solid) and $\Lambda_c=0.80$ (spherical type:dashed).
}
\end{figure}

Next, we discuss how the $^{28}$Si core structure in the $^{28}{\rm Si}+{\alpha}$ system 
is affected by the existence of an $\alpha$ cluster.
The $\alpha$ cluster at the surface of the $^{28}{\rm Si}$ core may affect the feature of the
$^{28}{\rm Si}$ core because of the nuclear and Coulomb interactions and also Pauli blocking effect.
To discuss features of the $^{28}{\rm Si}$ core with an $\alpha$ cluster at a certain distance $R$ from the core, 
we fix the parameter $\lambda_\alpha=0$ to assume the $\alpha$ cluster without the breaking,
and consider the $7\alpha$-breaking in the $^{28}$Si core and also the orientation of the 
oblate-type $^{28}$Si core in the $^{28}{\rm Si}+{\alpha}$ system.
Namely, we analyze the energy expectation value 
of the parity-projected state before the total-angular-momentum projection, 
\begin{eqnarray}
 E^+_{^{28}{\rm Si}+\alpha}(R,\theta,\lambda_{\alpha},\Lambda_c)
 = \frac{\braket{\Phi^+_{^{28}{\rm Si}+\alpha}(R,\theta,\lambda_{\alpha},\Lambda_c)|\hat{H}|\Phi^+_{^{28}{\rm Si}+\alpha}(R,\theta,\lambda_{\alpha},\Lambda_c)}}
 {{\braket{\Phi^+_{^{28}{\rm Si}+\alpha}(R,\theta,\lambda_{\alpha},\Lambda_c)|\Phi^+_{^{28}{\rm Si}+\alpha}(R,\theta,\lambda_{\alpha},\Lambda_c)}}},\\
\Phi^+_{^{28}{\rm Si}+\alpha}(R,\theta,\lambda_{\alpha},\Lambda_c)=\hat{P}^+\Phi_{^{28}{\rm Si}+\alpha}(R,\theta,\lambda_{\alpha},\Lambda_c),
\label{eq:EEV}
\end{eqnarray}
with $\lambda_\alpha=0$. 

Figure \ref{fig:proc_fig2} shows the  $^{28}{\rm Si}+\alpha$ energies for the oblate-type $^{28}$Si core 
($\Lambda_c=0.38$) and the spherical-type $^{28}$Si core ($\Lambda_c=0.80$) set at the orientation $\theta=0^{\circ}$.
The energies are plotted as functions of the inter-cluster distance $R$. 
It is found that, in the $R=8$ fm region, 
energies of the two cases ($\Lambda_c=0.38$ and $0.80$) almost degenerate as expected from the
energy degeneracy in the isolate $^{28}$Si.
In the $2 < R < 5$ fm region, the energy for the oblate core is lower than 
that for the spherical core  indicating that, when an $\alpha$ cluster exists at the surface,
the oblate-type $^{28}$Si core is energetically favored than the spherical-type
because of the smaller overlap, {\it i.e.}, the weaker Pauli blocking of nucleons between the $\alpha$ cluster and the core for the oblate core at $\theta=0^{\circ}$ than in the spherical core case.

\begin{figure}[h]
\begin{center}
\includegraphics[clip,width=9.0cm]{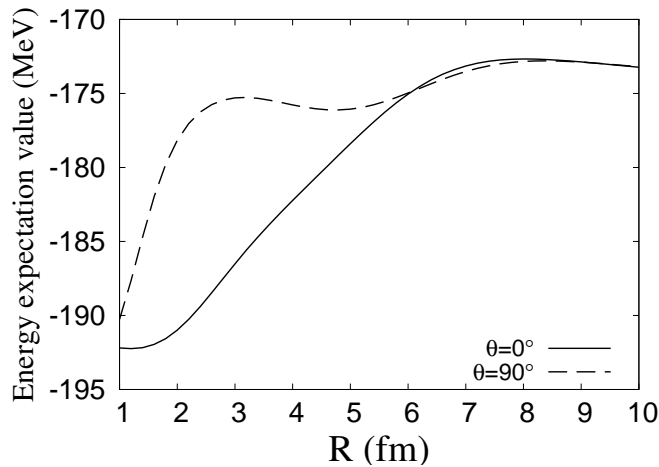}
\end{center}
\caption{\label{fig:S32_theta}
Energy expectation value of $^{28}{\rm Si}+\alpha$ system ($E^+_{^{28}{\rm Si}+\alpha}(R,\theta,\lambda_{\alpha}=0,\Lambda_c=0.38)$) as a function of the inter-cluster distance $R$.  The rotation angle $\theta$ of the oblate core is fixed to be $\theta=0^{\circ}$ (solid) and $\theta=90^{\circ}$ (dashed).
} 
\end{figure}

To see the $\theta$ dependence of the $^{28}{\rm Si}+\alpha$ energy,
we plot the energy expectation value 
$E^+_{^{28}{\rm Si}+\alpha}(R,\theta,\lambda_{\alpha}=0,\Lambda_c=0.38)$ of the 
oblate-type $^{28}$Si core oriented at 
$\theta=0^{\circ}$ and $90^\circ$ in Fig. \ref{fig:S32_theta}. 
In the small $R$ region ($R<5$ fm), the $\theta=0^{\circ}$ oriented core is favored 
because of the weaker Pauli blocking than the $\theta=90^{\circ}$ oriented core.
On the other hand, the energy does not depend on the core orientation in the large $R$ region,
in which the rotational symmetry of the $^{28}$Si core is restored.
In the $6<R<8$ fm region around the barrier, the $\theta=90^{\circ}$ oriented core gains slightly larger 
potential energy than the $\theta=0^{\circ}$ core but the energy difference is minor.

\subsection{$\alpha$-cluster breaking}
\label{alpha-cluster_breaking}

\begin{figure}[h]
\begin{center}
\includegraphics[clip,width=9.0cm]{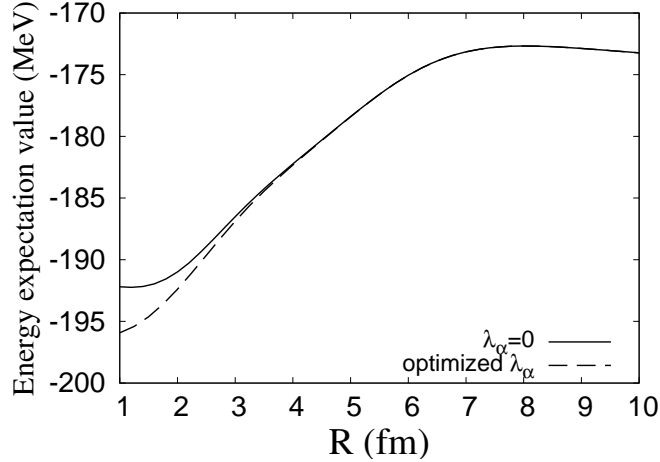}
\end{center}
\caption{\label{fig:alpha_breaking_s}
Energy expectation value of $^{28}{\rm Si}+\alpha$ system ($E^+_{^{28}{\rm Si}+\alpha}(R,\theta=0^{\circ},\lambda_{\alpha},\Lambda_c=0.38)$)
for the optimized $\lambda_{\alpha}$ as a function of the inter-cluster distance $R$ (dashed).
The energy for $\lambda_{\alpha}=0$ is also shown for comparison (solid).
} 
\end{figure}

We analyze the $\lambda_{\alpha}$ dependence of the energy expectation value of the $^{28}{\rm Si}+\alpha$ system
to see the $\alpha$-cluster breaking effect on the $^{28}{\rm Si}+\alpha$ system.
Figure \ref{fig:alpha_breaking_s} shows, 
the energy $E_{^{28}{\rm Si}+\alpha}(R,\theta=0^{\circ},\lambda_{\alpha},\Lambda_c=0.38)$
with the $\alpha$-cluster breaking, namely, $\lambda_\alpha$ optimized at each distance $R$,
compared with the energy for  $\lambda_{\alpha}=0$ without the $\alpha$-cluster breaking.
The energy gain by the $\alpha$-cluster breaking is very small except for the $R<3$ fm region.
This results indicates that the $\alpha$-cluster breaking in the $^{28}{\rm Si}+\alpha$ system is minor in the $\alpha$-cluster excited states having large amplitudes of the $\alpha$ cluster at the surface region ($4 < R <6$ fm).
Therefore, we ignore the $\alpha$-cluster breaking effect in the GCM calculation discussed in the next section for 
simplicity. 

In the $R<2$ fm region, the finite $\lambda_\alpha$ gives some energy gain to the
$^{28}{\rm Si}+\alpha$ system, but it is not appropriate to regard it as the $\alpha$-cluster breaking because the $\alpha$-cluster gets into in the inner region of the core and the $^{28}{\rm Si}+\alpha$ picture breaks down in this region.  
More details of the $\alpha$-cluster breaking in the $^{28}{\rm Si}+\alpha$ system are discussed later.

\subsection{GCM calculation}

We superpose $^{28}{\rm Si}+\alpha$ wave functions and obtain the ground and excited states of $^{32}$S with the full-GCM calculation described in Sec. \ref{FW:GCM}.

The calculated value of the $^{32}$S binding energy is $205.71$ MeV which underestimates the experimental binding energy ($271.78$ MeV),
whereas that of the $\alpha$-separation energy is $13.8$ MeV which overestimates the experimental value ($6.95$ MeV).
We can adjust the interaction parameter $M$ of the Volkov force to reproduce either the binding energy or the $\alpha$-separation energy,
but it is difficult to reproduce both data within the present two-body effective interaction. 
At the end of this section, 
we show energy levels calculated by using modified interaction parameters 
to see the interaction dependence of the result.

\begin{figure}[h]
\begin{center}
\includegraphics[clip,width=9.0cm]{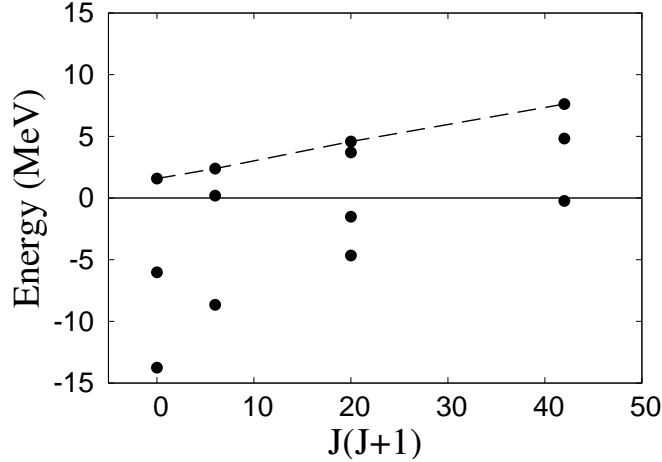}
\end{center}
\caption{\label{fig:GCM_out}
The energy levels of $^{32}$S obtained by the full-GCM calculation.
The energies are measured from the $^{28}{\rm Si}+\alpha$ threshold.
The dashed line indicates the members of the $\alpha$-cluster band. 
} 
\end{figure}

Figure \ref{fig:GCM_out} shows the energy levels $^{32}$S obtained by the full-GCM with the default 
interaction parameters. 
Energies measured from the $^{28}{\rm Si}+\alpha$ threshold energy are plotted as 
functions of  $J(J+1)$.
In the energy region near the $^{28}{\rm Si}+\alpha$ threshold, we obtain $J^{\pm}=0^+, 2^+, 4^+,$ and $6^+$ states 
having a remarkably developed $\alpha$-cluster structure. 
We assign these states as $\alpha$-cluster excited states belonging to an 
$\alpha$-cluster band. In Fig. \ref{fig:GCM_out}, the corresponding 
$\alpha$-cluster excited states are shown by circles connected by dashed lines. 
The bandhead $0^+$ state starts from $E_r=1.58$ MeV above the $^{28}{\rm Si}+\alpha$ threshold
and the rotational energy approximately follows the expression of the rigid rotor model: 
\begin{eqnarray}
 E_{rot} = \frac{\hbar^2}{2 \mathcal{J}}J(J+1),
\end{eqnarray}
with the rotational constant $ k = {\hbar^2}/{2 \mathcal{J}} = 145$ keV up to the $6^+$ state. 
We do not obtain an $\alpha$-cluster excited state with $J^{\pm}=8^+$. 
We also obtain other excited states lower than the $\alpha$-cluster excited states, 
but their energies change with the increase the number of bases 
and we can not obtain converged energies.  
This means that the present model space of the extended $^{28}{\rm Si}+\alpha$ cluster model
is not sufficient to describe non-cluster states of $^{32}$S in the low energy region.
On the other hand, we obtain good convergence for the energies of the ground state and 
the $\alpha$-cluster excited states with respect to the increase of the number of bases.  

\begin{figure}[h]
\begin{center}
\includegraphics[clip,width=14.0cm]{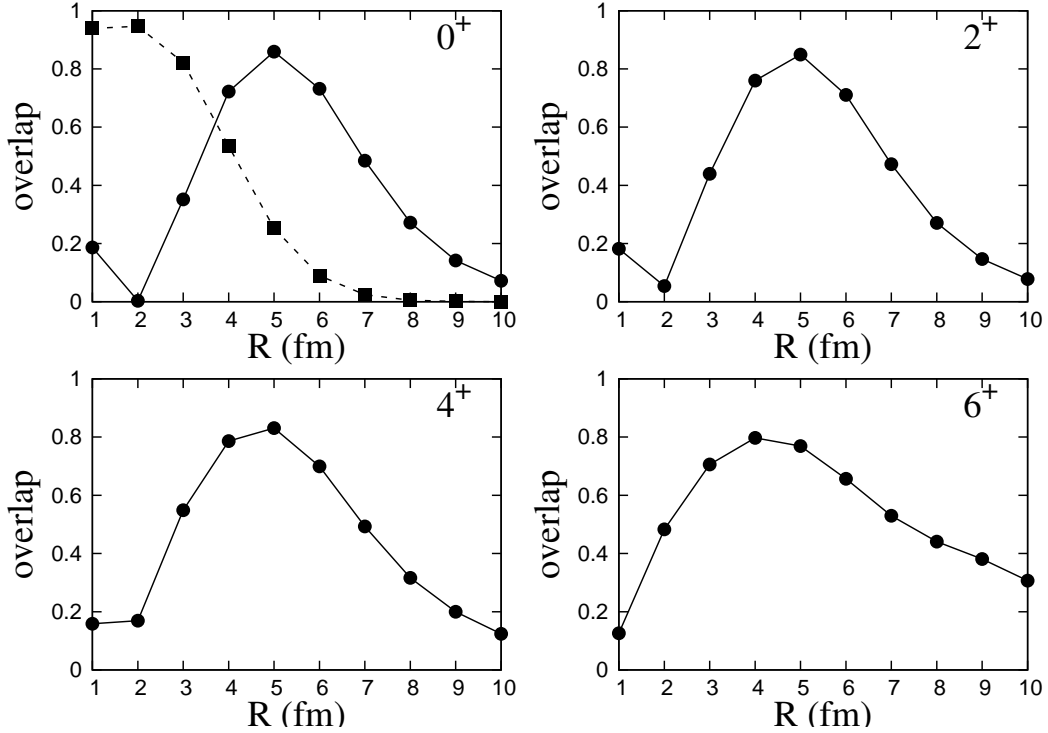}
\end{center}
\caption{\label{fig:amp1}
The overlap $f^{J^+_n}_{^{28}{\rm Si}+\alpha}(R)$ for the $\alpha$-cluster excited states with $J^{\pm}=0^+, 2^+, 4^+,$ and $ 6^+$ (filled circles).
The overlap for the ground state is also shown by filled squares in the upper left panel.
} 
\end{figure}

We show the overlap 
$f^{J^{\pm}_n}_{^{28}{\rm Si}+\alpha}(R)$
defined in Eq. (\ref{eq:overlap_full}) between the full-GCM wave function 
$\Psi^{J^{\pm}_n}_{^{28}{\rm Si}+\alpha}$
and the $R$-fixed frozen core wave function $\Phi^{J^{\pm}}_{^{28}{\rm Si}(0^+_{g.s.})+\alpha} (R)$
for the $\alpha$-cluster excited states ($J^{\pm}=0^+, 2^+, 4^+, 6^+$)
in Fig. \ref{fig:amp1}. 
We also show the overlap for the ground state.
The overlap $f^{J^{\pm}_n}_{^{28}{\rm Si}+\alpha}(R)$
indicates the $\alpha$-cluster amplitude at $R$ in the $L=J$ orbit around the $^{28}$Si ground state.
It is found that the ground state has no developed $\alpha$ cluster in the large $R$ region. 
In contrast to the ground state, the $\alpha$-cluster excited states show the developed $\alpha$-cluster in the large $R$ region: the $0^+, 2^+$, and $4^+$ states have large amplitudes in the $R\sim5$ fm region whereas the $6^+$ state has the peak at $R=4$ fm with a long tail in the large $R$ region.

\begin{table}[htb]
\caption{ 
The reduced $\alpha$-decay widths $\theta^2_\alpha(a)$ at the channel radii $a=6$ and $a=7$ fm and the partial $\alpha$-decay widths $\Gamma_\alpha$ for the $^{28}{\rm Si}(0^+_{g.s.})+\alpha$ channel in the $l=J$-wave.
The results calculated with $M=0.67$ are shown.
The $\alpha$-decay energies used in the calculation of $\Gamma_\alpha$ are the calculated values starting from $E_r=1.6$ MeV, and those shifted by 2.3 MeV to adjust the bandhead energy to the experimental value $E_r=3.9$ MeV reported in Ref.~ \cite{Lonnroth2011}.
}
  \label{table:alpha-decay}
  \centering
  \begin{tabular*}{15cm}{@{\extracolsep{\fill}} c|ccccc}
 \hline
  \multicolumn{6}{c}{Calculate ($M=0.67$)}  \\
  \hline
   State  & $E_r$ (MeV) & \multicolumn{2}{c}{$\Gamma_\alpha(a)$ (MeV)} 
     & \multicolumn{2}{c}{$\theta_\alpha^2(a)$} \\ 
   $J^{\pm}$ & Cal. & $a=6$ & $a=7$ & $a=6$ & $a=7$  \\ \hline
   $0^+$  & 1.6 & 9.9$\times 10^{-8}$ & 2.3$\times 10^{-7}$ & 0.32 & 0.16 \\ 
   $2^+$  & 2.4 & 2.3$\times 10^{-5}$ & 5.3$\times 10^{-5}$ & 0.30 & 0.16 \\ 
   $4^+$  & 4.6 & 0.0062 \ \, \, & 0.014 \, \ \ \,          & 0.29 & 0.17 \\ 
   $6^+$  & 7.6 & 0.039 \, \ \, \, & 0.098 \, \ \ \,        & 0.26 & 0.20 \\
  \hline
  \multicolumn{6}{l}{}  \\
  \cline{1-4}
  \multicolumn{4}{c}{Shifted}  \\
  \cline{1-4}
   State  & $E_r$ (MeV) & \multicolumn{2}{c}{$\Gamma_\alpha(a)$ (MeV)}  \\ 
   $J^{\pm}$ & Shifted & $a=6$ & $a=7$ \\ \cline{1-4}
   $0^+$  & 3.9 & 0.033 & 0.039      \\ 
   $2^+$  & 4.7 & 0.060 & 0.068      \\ 
   $4^+$  & 6.9 & 0.17 \, & 0.19 \,  \\ 
   $6^+$  & 9.9 & 0.23 \, & 0.34 \,  \\ 
  \cline{1-4}
  \end{tabular*}
\end{table}

We estimate the $\alpha$-decay widths of the $\alpha$-cluster excited states 
using the overlap $f^{J^{\pm}_n}_{^{28}{\rm Si}+\alpha}(R)$ defined
in Eq. (\ref{eq:overlap_full})  with the approximation method in Ref. \cite{Enyo2014}.
Following the method in Ref. \cite{Enyo2014}, the (dimensionless) reduced $\alpha$ width $\theta^2_\alpha(a)$ at the channel radius $a$ is approximately evaluated by the overlap as, 
\begin{eqnarray}
\theta^2_\alpha(a) &\approx& \frac{a}{3} \sqrt{\frac{\gamma}{2\pi}} 
\left( f^{J^{\pm}_n}_{^{28}{\rm Si}+\alpha}(a) \right)^2, \\
\gamma &=& \frac{A_1A_2}{A} \nu,
\end{eqnarray}
where $A$, $A_1$, and $A_2$ are the mass numbers of $^{32}$S, $^{28}$Si, and $\alpha$ cluster, respectively.
Using $\theta^2_\alpha(a)$, we calculated the partial $\alpha$-decay width $\Gamma_\alpha$ of the $^{28}{\rm Si}(0^+_{g.s.})+\alpha$ channel in the $L$-wave ($L=J$) as,
\begin{eqnarray}
\Gamma_\alpha &=& 2 P_L(a) \theta^2_\alpha(a) \gamma_{\rm w}^2(a), \\
P_L(a)        &=& \frac{ka}{F_L^2(ka)+G_L^2(ka)},
\end{eqnarray}
where $F_L$ and $G_L$ are the regular and irregular Coulomb functions, respectively, $\gamma_{\rm w}^2$ is the Wigner limit of the reduced $\alpha$-width $\gamma_{\rm w}^2=3\hbar^2/2\mu a^2$, $\mu$ is the reduced mass, and 
$k=\sqrt{2\mu E_r}/\hbar$. 
The calculated $\theta^2_\alpha(a)$ and $\Gamma_\alpha$ of the $\alpha$-cluster band in $^{32}$S are shown in Table \ref{table:alpha-decay}.
At $a=6$ fm, the reduced $\alpha$ widths are significant as $\theta^2_\alpha(a)=0.26 \sim 0.32$ reflecting the spatially developed cluster structure in this band. 
For the $\alpha$-decay widths, we calculate $\Gamma_\alpha$ in two cases of the bandhead energy considering the ambiguity of the predicted bandhead energy because the $\alpha$-decay width is quite sensitive to the $\alpha$-decay energy.
In the first case, we use the energies obtained in the present calculation, in which the bandhead energy is $E_r=1.6$ MeV.  In the second case, we shift the energies by $2.3$ MeV by hand to adjust the bandhead energy to the experimental value $E_r=3.9$ MeV reported by L\"onnroth {\it et al.} \cite{Lonnroth2011}.

Let us discuss comparison with the experimental reports of the $\alpha$-cluster excited states. 
In the experiment of elastic $^{28}{\rm Si}+\alpha$ scattering,
L\"onnroth {\it et al.} reported the $\alpha$-cluster excited band starting from the bandhead energy 
$E_r= 3.9\pm0.5$ MeV measured from the $^{28}{\rm Si}+\alpha$ threshold \cite{Lonnroth2011}. 
They evaluated the rotational constant $k= 122\sim 152$ keV from the averaged energies of the fragmented states.
In the experiment of $\alpha$ inelastic scattering on $^{32}{\rm S}$, Itoh {\it et al.} 
suggested candidates for two $\alpha$-cluster excited bands at bandhead energies $E_r=-0.4$ MeV and $E_r=0.9$ MeV with the rotational constants $k = 125$ keV and $k = 234$ keV, respectively \cite{Itoh2012}. 
The calculated bandhead energy $E_r=1.58$ MeV obtained with $M=0.67$ is
an intermediate value between those experimental reports.
The rotational constant $k = 145$ keV in the present result is within the range of the data reported by L\"onnroth {\it et al.} \cite{Lonnroth2011},
whereas it is slightly larger than the value $k = 125$ keV for the band at $E_r=-0.4$ MeV reported by Itoh {\it et al}.
Although the $\alpha$-cluster excited states observed by L\"onnroth {\it et al.} are fragmented, the fragmentation of the $\alpha$-cluster band is not found in the present result, because the present model space may be insufficient to describe the fragmentation.

\begin{figure}[t]
\begin{center}
\includegraphics[clip,width=9.0cm]{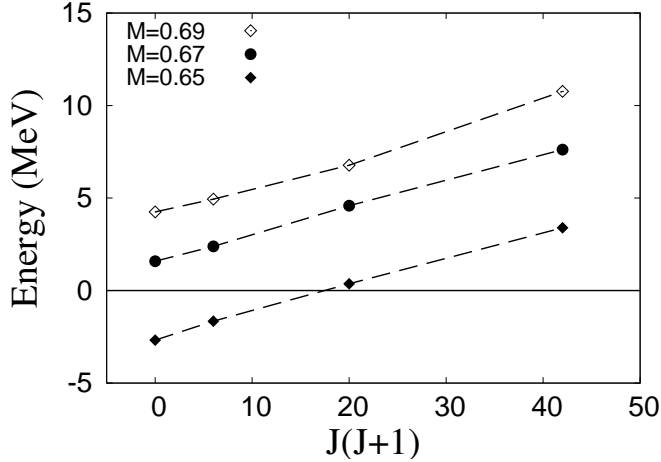}
\end{center}
\caption{\label{fig:GCM_M}
The energy levels of the $\alpha$-cluster band in $^{32}$S 
obtained by the full-GCM calculation using $M=0.69, 0.67,$ and $0.65$.
The energies measured from the $^{28}{\rm Si}+\alpha$ threshold are plotted.
} 
\end{figure}

As mentioned previously, it is difficult to reproduce experimental data of both the binding energy 
($271.8$ MeV) and the $\alpha$-separation energy ($6.95$ MeV) of $^{32}$S with the present effective interaction, and therefore, we can not quantitatively predict the energy positions of excited states. We here discuss the interaction parameter dependence of the energy position of the $\alpha$-cluster excited band.
We modify the Majorana parameter $M$ in the Volkov No.2 force from $M=0.67$ to $M=0.69$ which reproduces the $\alpha$-separation energy ($6.41$ MeV) but underestimates the binding energy of $^{32}$S ($172.2$ MeV).
We also use $M=0.65$ which gives the binding energy of $239.8$ MeV and the $\alpha$-separation energy of $21.74$ MeV.
In Fig.\ref{fig:GCM_M}, we show energy levels of the $\alpha$-cluster excited band obtained with $M=0.69, 0.67,$ and $0.65$.
The bandhead energy $E_r=4.25$ MeV and the rotational constant $k=149$ keV are obtained with $M=0.69$,
and $E_r=-2.68$ MeV and $k=146$ keV are obtained with $M=0.65$.
The bandhead energy depends on the interaction parameter and ranges from $E_r=-2.68$ MeV to $E_r=4.25$ MeV with these $M$ values. 
In contrast to the strong interaction dependence of the bandhead energy, 
the rotational constant is not sensitive to the interaction parameter in the present calculation.
Although it is difficult to quantitatively predict the handhead energy in the present calculation,
we  can say that the $\alpha$-cluster excited states appear near the $^{28}$Si+$\alpha$ threshold and construct the rotational band up to the $J^\pi=6^+$ state with the rotational constant $k=140 \sim 150$ keV.

%% file: DISCUSSION.tex

\section{DISCUSSION}
\label{sec:discussion}

\subsection{Core rotation and shape mixing effects}

\begin{figure}[h]
\begin{center}
\includegraphics[clip,width=9.0cm]{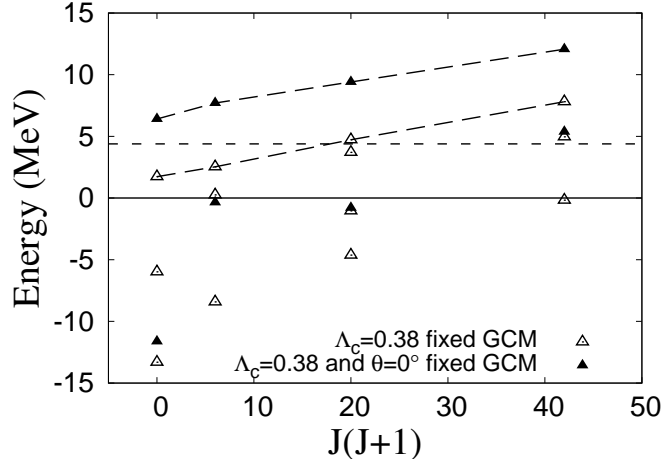}
\end{center}
\caption{\label{fig:change1}
The energy levels of $^{32}$S
 obtained by the $\Lambda_c=0.38$ and $\theta=0^{\circ}$ fixed GCM calculation (filled triangle) and the $\Lambda_c=0.38$ fixed GCM calculation (open triangle).
The energies are measured from the $^{28}{\rm Si}+\alpha$ threshold (solid line), and the $^{28}{\rm Si}(\Lambda_c=0.38)+\alpha$ threshold is plotted by the dotted line.
The $\alpha$-cluster bands are connected by dashed lines.
} 
\end{figure}

\begin{figure}[h]
\begin{center}
\includegraphics[clip,width=9.0cm]{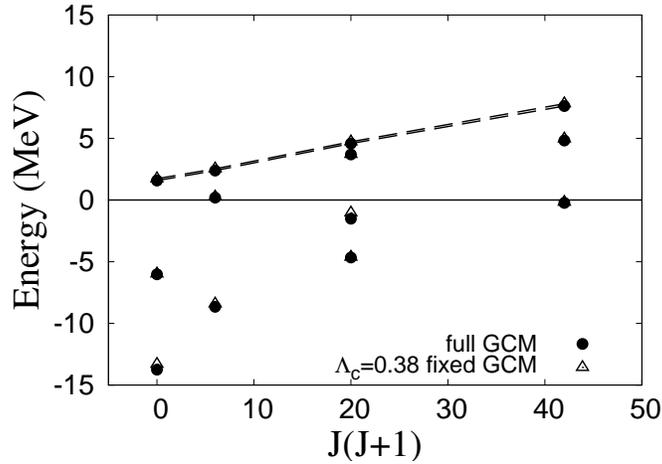}
\end{center}
\caption{\label{fig:change2}
The energy levels of $^{32}$S
obtained by the $\Lambda_c=0.38$ fixed GCM calculation and full-GCM calculation (filled circle).
The energies are measured from the $^{28}{\rm Si}+\alpha$ threshold (solid line).
The $\alpha$-cluster bands are connected by dashed lines.
} 
\end{figure}

In the full-GCM calculation, we take into account the core rotation and the oblate-spherical mixing as well as the inter-cluster motion
by superposing the parity and total-angular-momentum projected $^{28}$Si+$\alpha$ wave functions
with $R$, $\theta$, and $\Lambda_c$.
For the inter-cluster distance, $R=1,2,\cdots,10$ fm are used.
For the rotation angle of the $^{28}$Si core, $\theta=0^{\circ},
30^{\circ}, 60^{\circ}, 90^{\circ}$ are used for the oblate core ($\Lambda_c=0.38$),
and $\theta$ is fixed to be $\theta=0^{\circ}$ for the spherical core
($\Lambda_c=0.80$).
Here, we perform GCM calculations with reduced basis wave functions to 
discuss how the core rotation and the oblate-spherical mixing affect 
the $\alpha$-cluster excited states.

We discuss the core rotation effect on the energy spectra.
In Fig. \ref{fig:change1}, we compare the energy spectra obtained by 
the GCM calculations of the oblate core with and without the core rotation.
The former is calculated by superposing 
$^{28}$Si+$\alpha$ wave functions with $R=1,\cdots,10$ fm for the 
$\Lambda_c=0.38$ core at $\theta=(0^{\circ}, 30^{\circ}, 60^{\circ}, 90^{\circ})$, and
the latter is calculated by those with $R=1,\cdots,10$ fm for the $\Lambda_c=0.38$ core at the fixed angle $\theta=0^{\circ}$.
The mixing of the spherical core ($\Lambda_c=0.80$) is omitted in this analysis for simplicity.
As the result, the energy reduction by the core rotation is remarkable for the 
$\alpha$-cluster excited states. The band energy is reduced by about $5$ MeV, 
which is almost consistent with $4.4$ MeV reduction of the 
$^{28}{\rm Si}(\Lambda_c=0.38)+\alpha$ threshold caused by the $0^+$ projection of $^{28}{\rm Si}$.
It indicates that, in the $\alpha$-cluster excited states, the $\alpha$ cluster spatially develops 
and does not disturb the oblate core rotation. 

In Fig.~\ref{fig:change2}, we show the energy spectra obtained by
the GCM calculation with full base wave functions and that without the 
spherical core ($\Lambda_c=0.80$) wave functions to see the effect of the oblate-spherical mixing.
The result shows that the spherical core mixing effect is minor.

\subsection{Analysis in the weak coupling picture: frozen core GCM calculation}
\label{28Si_core_excitation_effects}

\begin{figure}[h]
\begin{center}
\includegraphics[clip,width=9.0cm]{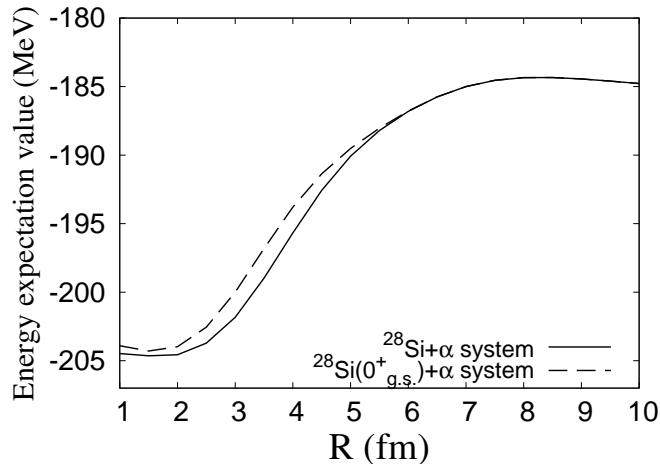}
\end{center}
\caption{\label{fig:EEV_AS}
Energy expectation values of the $R$-fixed $^{28}{\rm Si}+\alpha$ system and the $R$-fixed $^{28}{\rm Si}(0^+_{g.s.})+\alpha$ system are plotted.
The solid line is the $^{28}{\rm Si}+\alpha$ system and the dashed line is the $^{28}{\rm Si}(0^+_{g.s.})+\alpha$ system.
} 
\end{figure}

\begin{figure}[h]
\begin{center}
\includegraphics[clip,width=9.0cm]{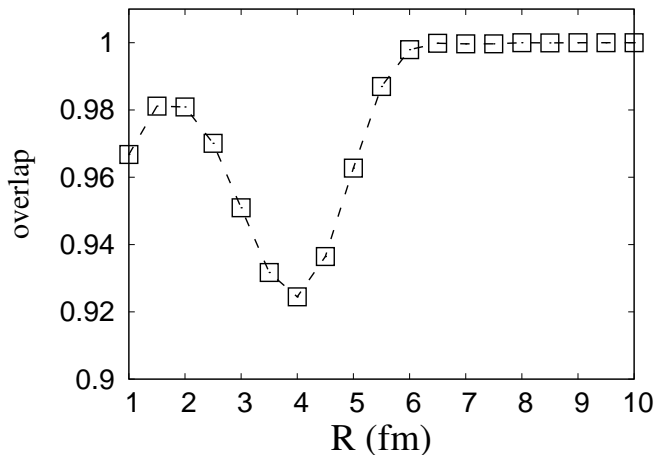}
\end{center}
\caption{\label{fig:overlap_AS}
The wave function overlap $f(R)$ between the $R$-fixed $^{28}{\rm Si}+\alpha$ system and
the $R$-fixed $^{28}{\rm Si}(0^+_{g.s.})+\alpha$ system defined in Eq. (\ref{eq:overlap}).
} 
\end{figure}

\begin{figure}[h]
\begin{center}
\includegraphics[clip,width=9.0cm]{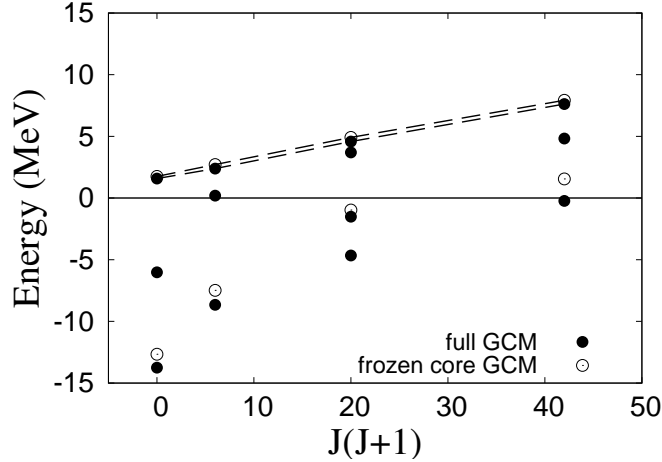}
\end{center}
\caption{\label{fig:GCM_AS}
The energy levels of $^{32}$S obtained by the frozen core GCM calculation (open circle) and the full-GCM calculation (filled circle).
The energies are measured from the $^{28}{\rm Si}+\alpha$ threshold (solid line).
The $\alpha$-cluster bands are connected by dashed lines.
} 
\end{figure}

\begin{figure}[h]
\begin{center}
\includegraphics[clip,width=14.0cm]{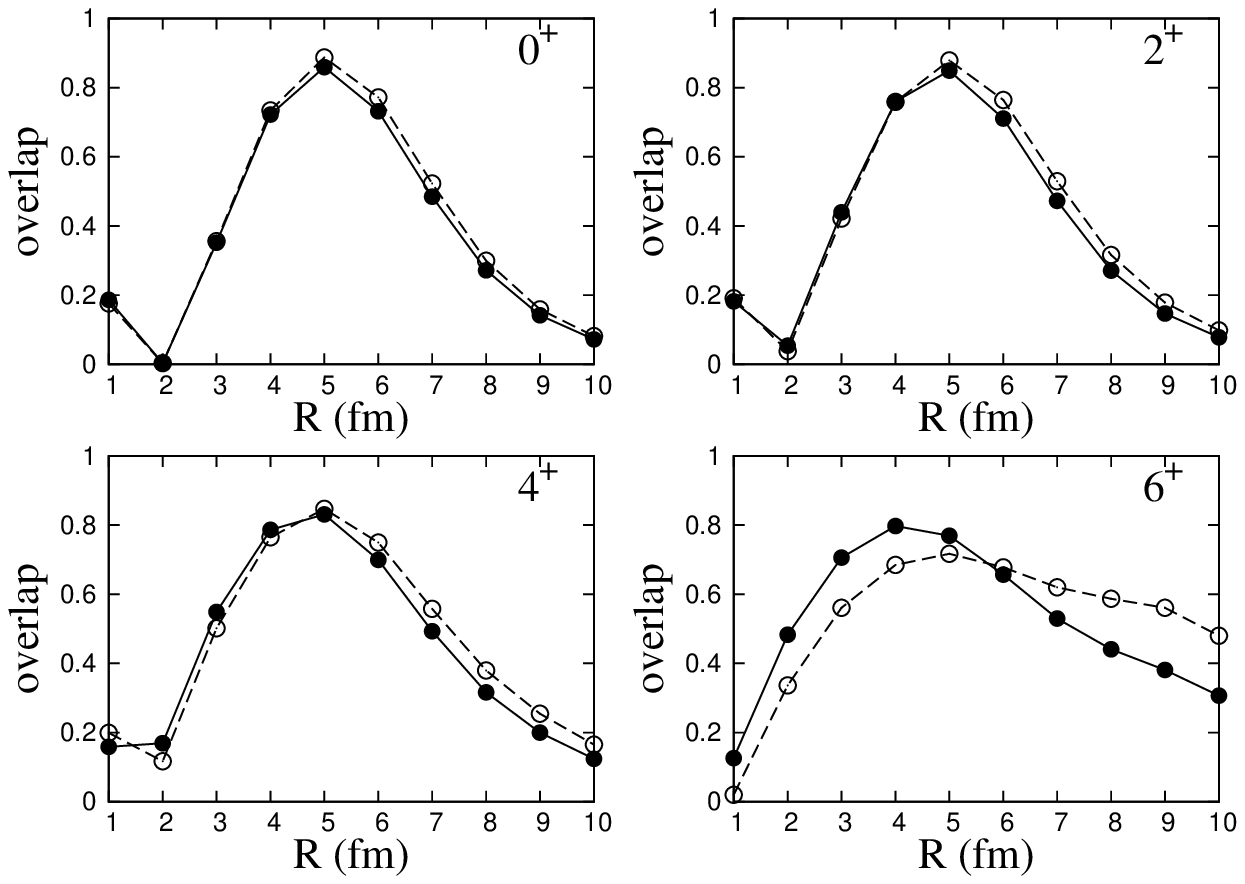}
\end{center}
\caption{\label{fig:overlap_frozen}
The overlaps between the full-GCM wave function $\Psi^{J^{\pm}_n}_{^{28}{\rm Si}+\alpha}$ and the $R$-fixed frozen core wave function $\Phi^{J^{\pm}}_{^{28}{\rm Si}(0^+_{g.s.})}(R)$ (filled circle),
and those between the frozen core wave function $\Psi^{J^{\pm}_n}_{^{28}{\rm Si}(0^+_{g.s.})+\alpha}$ and the $R$-fixed frozen core wave function $\Phi^{J^{\pm}}_{^{28}{\rm Si}(0^+_{g.s.})}(R)$ (open circle) of the $\alpha$-cluster band.
} 
\end{figure}

In the asymptotic region at a large inter-cluster distance $R$, the $^{28}{\rm Si}$ core should be the ground state of the isolate $^{28}{\rm Si}(0^+_{g.s.})$.
As discussed previously, the $\alpha$-cluster excited states contain dominantly the $^{28}{\rm Si}(0^+_{g.s.})$+$\alpha$ component.
Therefore, it is expected that the frozen core GCM calculation with the inert $^{28}{\rm Si}(0^+_{g.s.})$ core assumption can be a leading order approximation at least for the $\alpha$-cluster excited states.
The frozen core GCM calculation is the extreme case of the weak coupling and it is different from the adiabatic picture of the strong coupling.
In the previous section, we start from the strong coupling picture, in which the deformed $^{28}$Si core is located at a fixed orientation, and then consider the rotation and shape mixing effects on the $\alpha$-cluster excited states obtained by the full GCM calculation.
In this section, we discuss the features of the $\alpha$-cluster excited states from the weak coupling picture.
Namely, we start from the frozen core $^{28}{\rm Si}(0^+_{g.s.})$+$\alpha$ states, and then consider the effect of the core excitations, in particular, the rotational excitation from the $^{28}{\rm Si}(0^+_{g.s.})$.
Note that the core excitations taken into account in the present model are the rotational excitation such as $^{28}{\rm Si}(2^+)$ and also the change of the oblate-spherical mixing (shape mixing) from the $^{28}{\rm Si}(0^+_{g.s.})$.

After comparing the properties of the $R$-fixed $^{28}$Si+$\alpha$ wave function between the optimized $^{28}$Si core and the inert $^{28}{\rm Si}(0^+_{g.s.})$ core cases, we compare the result of the frozen core GCM calculation with that of the full GCM calculation containing the rotational and shape-mixing excitations from the $^{28}{\rm Si}(0^+_{g.s.})$ core.

For a certain inter-cluster distance $R$, we define the $R$-fixed frozen core wave function $\Phi^{J^{\pm}}_{^{28}{\rm Si}(0^+_{g.s.})+\alpha} (R)$ in Eq.~\eqref{P_si0+a(R)}, and also the $R$-fixed $^{28}{\rm Si}+\alpha$ wave function $\Phi^{J^{\pm}}_{^{28}{\rm Si}'+\alpha} (R)$ in Eq.~\eqref{P_si+a(R)}, where the $^{28}{\rm Si}$ core wave function is optimized so as to minimize the energy expectation value of the $R$-fixed $^{28}{\rm Si}+\alpha$ wave function.
We here consider $0^+$ projected wave functions. 
In the asymptotic region at a large inter-cluster distance $R$, $\Phi^{J^{\pm}}_{^{28}{\rm Si}'+\alpha} (R)$ equals to $\Phi^{J^{\pm}}_{^{28}{\rm Si}(0^+_{g.s.})+\alpha} (R)$. 
On the other hand $\Phi^{J^{\pm}}_{^{28}{\rm Si}'+\alpha} (R)$ may deviate from $\Phi^{J^{\pm}}_{^{28}{\rm Si}(0^+_{g.s.})+\alpha} (R)$, in the short inter-cluster distance region, in which the core excitation from the $^{28}{\rm Si}(0^+_{g.s.})$ occurs because of the existence of the $\alpha$ cluster to gain the total energy.

We plot the energy expectation values of the $R$-fixed frozen core wave function
and the $R$-fixed $^{28}{\rm Si}+\alpha$ wave function in Fig. \ref{fig:EEV_AS}. 
In Fig. \ref{fig:overlap_AS}, we show the overlap 
between the $R$-fixed $^{28}{\rm Si}+\alpha$ wave function and the frozen core wave function,
\begin{eqnarray}
f(R) = \left| \braket{ \Psi^{0^+}_{^{28}{\rm Si}(0^+_{g.s.})+\alpha}(R)| \Psi^{0^+}_{^{28}{\rm Si}'+\alpha}(R)  } \right|,
\label{eq:overlap}
\end{eqnarray}
which is reduced from $1$ by the core excitation.
It is found that the core excitation from the $^{28}{\rm Si}(0^+_{g.s.})$ occurs in the  $R<6$ fm region
and it reduces the energy of the total system $^{32}$S in $R\le 5$ fm.
These results indicate that the $R>6$ fm region is understood as the ideal weak coupling regime of $^{28}{\rm Si}(0^+_{g.s.})$+$\alpha$, whereas the rotational and shape-mixing excitations of the $^{28}{\rm Si}$ core occur in the $R<6$ fm region.

Next, we compare the frozen core GCM calculation given by Eq. (\ref{eq:wf_frozen_core}) 
with the full-GCM calculation to see the core excitation effects in particular on the $\alpha$-cluster band. 
Figure \ref{fig:GCM_AS} shows the energy spectra obtained by the full-GCM and the frozen core GCM calculations.
The energy of the ground state decreases by about $1$ MeV from the frozen core GCM to the full-GCM calculation.
The energy of the $\alpha$-cluster band also shifts down slightly because of the core excitation effect.

In Fig. \ref{fig:overlap_frozen}, we compare the overlap 
$f^{J^{\pm}_n}_{^{28}{\rm Si}(0^+_{g.s.})+\alpha}(R)$
(Eq. \ref{eq:overlap_frozen})
for the frozen core GCM and
$f^{J^{\pm}_n}_{^{28}{\rm Si}+\alpha}(R)$
(Eq. \ref{eq:overlap_full})
for the full-GCM.
Compared with the $\alpha$-cluster amplitudes for the frozen core GCM calculation, 
those in the full-GCM calculation tend to be slightly suppressed in the outer region ($R\ge5$ fm). 
It indicates that the $\alpha$ cluster is attracted toward the inner region because of the $^{28}$Si core excitation such as deformation and rotation,
which gives additional attraction in the $R<5$ fm region as discussed previously.
In other words, the core excitation plays a role to stabilize the $\alpha$-cluster excited states.

\subsection{$\alpha$-cluster breaking at the nuclear surface}

\begin{figure}[h]
\begin{center}
\includegraphics[clip,width=9.0cm]{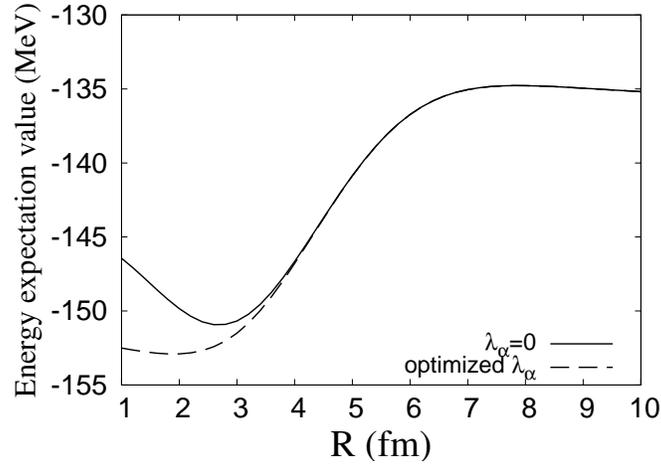}
\end{center}
\caption{\label{fig:alpha_breaking_ne}
Energy expectation value of the $^{16}{\rm O}+\alpha$ system for the optimized $\lambda_{\alpha}$ as a function of the inter-cluster distance $R$ (dashed line). 
The energy for $\lambda_{\alpha}=0$ is also shown for comparison (solid line).
} 
\end{figure}

\begin{figure}[h]
\begin{center}
\includegraphics[clip,width=9.0cm]{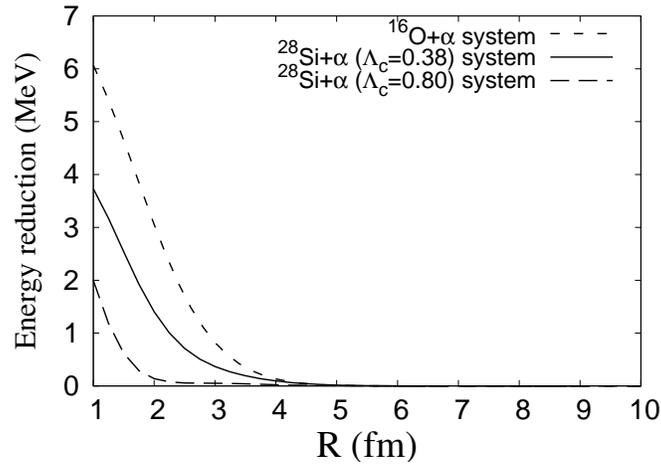}
\end{center}
\caption{\label{fig:energy_reductions}
The energy difference between the cases of the optimized $\lambda_\alpha$ and the fixed $\lambda_{\alpha}=0$ for the $^{16}{\rm O}+\alpha$, $^{28}{\rm Si}+\alpha$ ($\Lambda_c=0.38$), and $^{28}{\rm Si}+\alpha$ ($\Lambda_c=0.80$) systems.
}
\end{figure}

\begin{figure}[h]
\begin{center}
\includegraphics[clip,width=9.0cm]{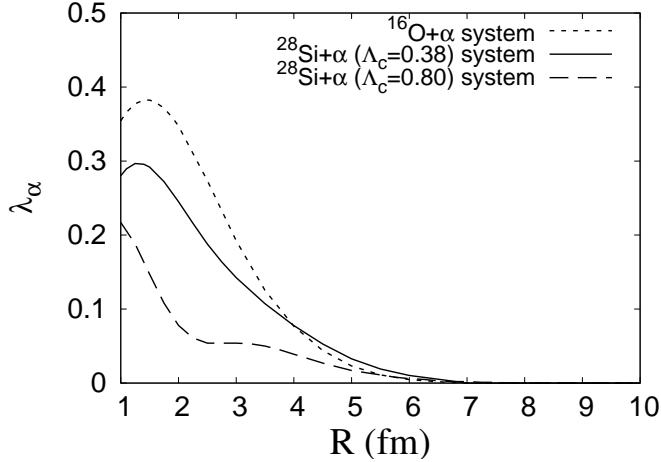}
\end{center}
\caption{\label{fig:opt_alpha}
The $\alpha$-cluster breaking parameter $\lambda_{\alpha}$ optimized to minimize the energies of the $^{16}{\rm O}+\alpha$, $^{28}{\rm Si}+\alpha$ ($\Lambda_c=0.38$), and $^{28}{\rm Si}+\alpha$ ($\Lambda_c=0.80$) systems.
} 
\end{figure}

As mentioned in Sec. \ref{alpha-cluster_breaking}, the $\alpha$-cluster breaking 
around the $^{28}$Si core is minor in the surface region.
We here discuss details of the $\alpha$-cluster breaking around the $^{28}$Si core 
in the $^{28}$Si+$\alpha$ system in comparison with that around the $^{16}$O core
in the $^{16}$O+$\alpha$ system to clarify the core dependence of the $\alpha$-cluster breaking. 

We perform a similar analysis of the $\alpha$-cluster breaking for the $^{16}$O+$\alpha$ system
by using the following $^{16}$O+$\alpha$ model wave function,
\begin{eqnarray}
 \Phi_{^{16}{\rm O}+\alpha}(R,\lambda_{\alpha})
&=& \mathcal{A} \left[ 
     \Phi_{\alpha'}        \left(\frac{4}{5}R{\bf e}_z,\lambda_{\alpha}\right) 
     \Phi_{^{16}{\rm O}} \left(-\frac{1}{5}R{\bf e}_z\right)
               \right],\\
\Phi_{^{16}{\rm O}}({\bf R})&=&\hat{T}({\bf R})\Phi_{^{16}{\rm O}},
\end{eqnarray}
where the $^{16}{\rm O}$ core wave function $\Phi_{^{16}{\rm O}}$ is written by a tetrahedron formed 4$\alpha$-clusters with the $\alpha$-$\alpha$ distance $0.5$ fm which is almost equivalent to the 
double closed $p$-shell configuration.
The width parameter is taken to be $\nu=0.195\ {\rm fm}^{-2}$.
The $\lambda_\alpha$ is optimized to minimize
the energy expectation value of the parity-projected $^{16}{\rm O}+\alpha$ wave function,
\begin{eqnarray}
 E^+_{^{16}{\rm O}+\alpha}(R,\lambda_{\alpha})
 &=& \frac{\braket{ P^+\Phi_{^{16}{\rm O}+\alpha}(R,\lambda_{\alpha})|\hat{H}|P^+\Phi_{^{16}{\rm O}+\alpha}(R,\lambda_{\alpha}) }}
          {\braket{ P^+\Phi_{^{16}{\rm O}+\alpha}(R,\lambda_{\alpha})        |P^+\Phi_{^{16}{\rm O}+\alpha}(R,\lambda_{\alpha}) }}.
\end{eqnarray}

Figure \ref{fig:alpha_breaking_ne} shows the energy of the $R$-fixed $^{16}$O+$\alpha$ wave function with the optimized $\lambda_\alpha$ (with the $\alpha$-cluster breaking) and that with the fixed $\lambda_\alpha=0$ (without the $\alpha$-cluster breaking).
As discussed previously, for the $^{28}$Si core case, the energy reduction by the $\alpha$-cluster breaking is found only in the very short distance region (see Fig. \ref{fig:alpha_breaking_s}), whereas there is almost no energy reduction
in the $R\ge 3$ fm region where the $\alpha$-cluster excited states have the $\alpha$-cluster amplitudes.
Differently from the $^{28}$Si+$\alpha$ system, in the $^{16}{\rm O}+\alpha$ system, the significant energy reduction by the $\alpha$-cluster breaking is found in a relatively wide $R$ region.
This energy reduction by the $\alpha$-cluster breaking shifts the energy minimum position to the short distance region, and it may give a significant effect to $\alpha$-cluster structure in the ground band of $^{20}$Ne as discussed in Ref. \cite{Itagaki2011}.
In Fig. \ref{fig:energy_reductions}, we show energy reductions by the $\alpha$ breaking, {\it i.e.}, the energy difference between the optimized $\lambda_\alpha$ and the fixed $\lambda_\alpha=0$ cases
for the $^{16}{\rm O}+\alpha$ and $^{28}{\rm Si}+\alpha$ ($\Lambda_c=0.38$) systems. 
We also show the energy reduction for the spherical $^{28}$Si core ($\Lambda_c=0.80$) case.
It is found that, the energy reduction of $^{28}{\rm Si}$+$\alpha$ ($\Lambda_c=0.38$) system is about a half of that of $^{16}{\rm O}+\alpha$ system in the $R=2 \sim 3$ fm region, and that of $^{28}{\rm Si}$+$\alpha$ ($\Lambda_c=0.80$) system is quite small. 
Thus the $\alpha$-cluster breaking gives energetically less important effects to the $^{28}{\rm Si}$+$\alpha$ system than to the $^{16}{\rm O}+\alpha$ system.

In Fig. \ref{fig:opt_alpha}, we compare the optimized values of the $\alpha$-breaking parameter $\lambda_\alpha$ for each system.
In both cases of the oblate and spherical $^{28}$Si cores, $\lambda_{\alpha}$ of the $^{28}{\rm Si}+\alpha$ system is smaller than that of the $^{16}{\rm O}+\alpha$ system at least in the $R<4$ fm region.
This indicates that, compared with the $^{16}{\rm O}+\alpha$ system, the $\alpha$-cluster breaking is relatively suppressed in $^{28}{\rm Si}+\alpha$, in particular, 
for the case of the spherical-type $^{28}$Si core $(\Lambda_c=0.80)$.

The $\alpha$-cluster breaking at the nuclear surface is caused mainly by the spin-orbit potential from the core nucleus, 
and therefore, it is naively expected that the $\alpha$-cluster breaking is likely to occur 
in heavier core systems because of the stronger core potential than light core systems.
The present result is opposite to this expectation. The reason is understood by the Pauli blocking effect from the $^{28}$Si core as follows. 
In general, in the $\alpha$-cluster breaking mechanism at the nuclear surface, 
4 nucleons in the broken $\alpha$ cluster favor to occupy the ls-favored orbits to gain the spin-orbit potential from the core rather than to form the ideal $(0s)^4$ $\alpha$-cluster.
However, in the $^{28}{\rm Si}+\alpha$ system, the ls-favored $0d_{5/2}$ orbits are occupied by
nucleons in the $^{28}$Si core, which block the $\alpha$-cluster breaking. 
The $0d_{5/2}$ orbits are fully blocked, in the $jj$-coupling limit $\Lambda_c=1$ for the sub-shell $0d_{5/2}$-closed $^{28}$Si core. Even though the $^{28}$Si core in the $^{28}$Si+$\alpha$ system is not in this limit,  
it has a finite $\Lambda_c$ and partially blocks the $0d_{5/2}$ orbits.
This picture can describe the suppression of the $\alpha$-cluster breaking at the surface of the $^{28}$Si core compared with that of the $^{16}$O core where $0d_{5/2}$ orbits are empty, and also the larger suppression for the spherical-type $(\Lambda_c=0.80)$ $^{28}$Si than that for the oblate-type ($\Lambda_c=0.38$) $^{28}$Si core.

%% file: CONCLUSION.tex

\section{CONCLUSION}
\label{sec:conclusion}

We investigated the $\alpha$-cluster excited states in $^{32}{\rm S}$. 
We proposed an extended model of the $^{28}$Si+$\alpha$ cluster model by taking into account the $^{28}$Si core deformation and rotation as well as the $\alpha$-cluster breaking.
The $^{28}$Si core is described by the extended 7$\alpha$-cluster model with the cluster breaking due to the spin-orbit interaction.

Applying the extended $^{28}$Si+$\alpha$ cluster model, we performed the GCM calculation and obtain the $\alpha$-cluster excited states near the $^{28}$Si+$\alpha$ threshold energy. 
These states construct the rotational band up to the $6^+$ state with the rotational constant $k=140\sim150$ keV.
We can not quantitatively predict the bandhead energy because of the ambiguity of the interaction parameters.  
The $\alpha$-cluster excited band obtained in the present work may correspond to one of the experimentally reported bands \cite{Lonnroth2011,Itoh2012}. 
The calculated rotational constant reasonably agrees to the value of the experimental band reported in Ref.~\cite{Lonnroth2011}.
Although the fragmentation of the $\alpha$-cluster excited states was observed in the experiment of Ref.~\cite{Lonnroth2011}, no fragmentation is found in the present calculation, maybe, because of the insufficient model space.

From the point of view of the strong coupling picture, we discussed the $^{28}$Si core deformation and rotation effects as well as the $\alpha$-cluster breaking one in the $\alpha$-cluster excited states.
It is found that the rotation of the oblately deformed $^{28}{\rm Si}$ core significantly reduces the excitation energies of the $\alpha$-cluster excited states, whereas the $\alpha$-cluster breaking gives only a minor effect.
We also analyzed the feature of the $\alpha$-cluster excited band from the weak coupling picture using the frozen core $^{28}{\rm Si}(0^+_{g.s.})+\alpha$ wave functions.
The $\alpha$-cluster excited states are found to have the dominant $^{28}{\rm Si}(0^+_{g.s.})+\alpha$ components.
The dimensionless reduced $\alpha$ widths estimated by the $^{28}{\rm Si}(0^+_{g.s.})+\alpha$ components are significantly large as $\theta^2_\alpha(a)=0.26 \sim 0.32$ at $a=6$ fm. We evaluated the partial $\alpha$-decay widths from the calculated values of $\theta^2_\alpha(a)$.
We also compared the result of the frozen core GCM calculation with that of the full GCM calculation, and found that the rotational excitation from the $^{28}{\rm Si}(0^+_{g.s.})$ plays an role to stabilize the $\alpha$-cluster excited states.

The present model is the extended $^{28}$Si+$\alpha$ cluster model, in which the cluster breaking due to the spin-orbit interaction and also the rotation of the deformed core are taken into account. The cluster breaking effect of the $^{28}$Si core part gives the large energy reduction ($18$ MeV) of the isolate $^{28}$Si from the 7$\alpha$-cluster model without the cluster breaking. This is an 
advantage over conventional cluster models using the Brink-Bloch $\alpha$-cluster model. 
Moreover, the rotation effect of the deformed core in $^{32}$S gives about 5 MeV reduction of the $\alpha$-cluster band energy from that obtained with the fixed core orientation. This indicates the importance of the angular momentum projection of the subsystem in the $\alpha$-cluster excited states having the deformed core. 

%% file: ACKNOWLEDGEMENTS.tex

\section{Acknowledgements}

The numerical calculations were carried out on SR16000 at YITP in Kyoto University.